\documentclass[%
 aip,jcp,
 amsmath,amssymb,
 reprint,
]{revtex4-2}

\usepackage[a-1b]{pdfx} 

\usepackage{graphicx}
\usepackage{dcolumn}
\usepackage{bm}

\usepackage[utf8]{inputenc}
\usepackage[T1]{fontenc}
\usepackage{mathptmx}
\usepackage{listings}
\usepackage{rotating} 

\usepackage{color}
\usepackage{dcolumn} 
\usepackage{bm} 
\usepackage{graphicx}
\usepackage{multirow} 
\usepackage{pifont} 
\usepackage{epsfig}
\usepackage{amsmath} 
\usepackage{subfigure}
\usepackage{float}
\usepackage{booktabs}
\usepackage{tabularx}
\usepackage{natbib}
\usepackage{gensymb}
\usepackage{rotating}
\usepackage{threeparttable}
\usepackage{comment}
\usepackage[normalem]{ulem}

\usepackage{pythonhighlight}

\begin{document}

\author{A. Eugene DePrince III}
\email{adeprince@fsu.edu}
\affiliation{
             Department of Chemistry and Biochemistry,
             Florida State University,
             Tallahassee, FL 32306-4390}

\title{Variational Determination of the Two-Electron Reduced Density Matrix: A Tutorial Review}

\begin{abstract}

The two-electron reduced density matrix (2RDM) carries enough information to evaluate the electronic energy of a many-electron system. The variational 2RDM (v2RDM) approach seeks to determine the 2RDM directly, without knowledge of the wave function, by minimizing this energy with respect to variations in the elements of the 2RDM, while also enforcing known $N$-representability conditions. 
In this tutorial review, we provide an overview of the theoretical underpinnings of the v2RDM approach and the $N$-representability constraints that are typically applied to the 2RDM. We also discuss the semidefinite programming (SDP) techniques used in v2RDM computations and provide enough Python code to develop a working v2RDM code that interfaces to the \texttt{libSDP} library of SDP solvers. 


\end{abstract}

\maketitle

\section{Introduction}

\label{SEC:INTRODUCTION}

The objective of supplanting the wavefunction with the two-electron reduced density matrix (2RDM) in practical quantum chemical calculations was dubbed “Coulson’s Challenge” by John Coleman, in reference to comments made by Charles Coulson in 1959.\cite{ColemanBook} The foundation for this challenge was laid almost two decades prior. In 1940,\cite{Husimi:1940:264} Husimi showed that the 2RDM carries sufficient information to define the energy for particles that interact via at most pairwise interactions. Later, L\"{o}din presented the same result and proposed that the 2RDM could be obtained directly by minimizing the energy with respect to variations in its elements.\cite{Lowdin:1955:1474}  Early calculations,\cite{Mayer:1955:1579} however, resulted in energies that were too low, in apparent violation of the variation theorem, because the space of 2RDMs over which the optimization was performed was larger than the space of 2RDMs that are derivable from an $N$-electron wave function or density matrix.\cite{Tregold1957_1421}  Coleman referred\cite{Coleman63_668} to those reduced density matrices (RDMs) that are derivable from an ensemble of $N$-electron density matrices as $N$-representable, and in 1963, Coleman solved the ensemble-state $N$-representability problem for the one-electron RDM (1RDM).\cite{Coleman63_668} Complete ensemble-state $N$-representability conditions for the 2RDM, however, remained elusive. In 1964, Garrod and Percus developed a set of necessary constraints on the 2RDM that were expressible in terms of the 2RDM (the P or D constraint), the two-hole RDM (the Q constraint) and the particle-hole RDM (the G constraint). Some numerical applications to nuclear structure followed,\cite{Rosina75_221,Rosina75_221,Garrod75_300} but it was not until the early 2000s that practical calculations that enforced the PQG (or DQG) conditions on molecular many-electron systems began to emerge.\cite{Fujisawa01_8282, Mazziotti02_062511} Around that time, higher-order constraints were derived\cite{Mazziotti02_062511, Percus04_2095, Erdahl01_042113} and numerical studies\cite{Mazziotti05_062503, Zhao08_164113, Mazziotti05_032510, Mazziotti06_144102, Mazziotti06_032501, Mazziotti07_024105} demonstrated that (partial) three-particle conditions are often sufficient to reach chemical accuracy. More recently, Mazziotti has put forward a formal solution to the ensemble-state $N$-representability problem for the 2RDM.\cite{Mazziotti12_263002, Mazziotti12_062507, Mazziotti23_153001} 

Given the apparent successes of v2RDM theory in the early 2000s, a great deal of effort has gone into the development of efficient numerical procedures for the variational optimization of the 2RDM,\cite{Mazziotti04_213001, Zhao07_553, Lewin06_064101, Zhao08_164113, Bultinck09_032508, DeBaerdemacker11_1235, Mazziotti11_083001, Mazziotti16_153001, DePrince19_6164, Mazziotti20_052819} and some codes have been applied to systems of more than 50 electrons distributed among more than 50 orbitals.\cite{Mazziotti08_134108, Mazziotti10_014104, DePrince16_423, DePrince16_2260, DePrince19_276, DePrince19_6164} Coupled to the fact that the v2RDM approach is naturally a multi-reference one, the ability to treat large numbers of electrons makes v2RDM attractive as a polynomially-scaling solver for large-scale approximate complete active space (CAS) configuration interaction (CASCI) or CAS self-consistent field (CASSCF) calculations.\cite{Mazziotti08_134108, Mazziotti11_5632, DePrince16_423, DePrince16_2260, DePrince19_276, DePrince19_6164} The v2RDM approach can also be used to obtain approximate 2RDMs corresponding to other multi-reference wave functions, namely, the doubly occupied configuration interaction (DOCI) wave function.\cite{VanNeck15_4064, Mazziotti17_084101, DeBaerdemacker18_024105, Lain18_194105, DePrince19_244121, Alcoba21_013110} In this case, due to the special structure of the DOCI RDMs, the application {\color{black} of} three-body $N$-representability conditions can be achieved at mean-field scaling [$\mathcal{O}(n^4)$, where $n$ is the number of correlated orbitals]. 

Despite these successes, v2RDM has remained a niche approach, for several reasons. First, v2RDM theory displays some unfortunate fundamental limitations\cite{Ayers09_5558, Bultinck10_114113, Cooper11_054115}that are similar to those encountered in density functional theory (the lack of a derivative discontinuity in the energy for fractionally charged systems, for example\cite{Yang08_792}). Second, publicly available implementations of the method are rare; we are aware of only three.  DePrince and coworkers have developed v2RDM solvers in the Q-Chem package\cite{Krylov21_084801} and in \texttt{hilbert},\cite{hilbert} which is a plugin to the \textsc{Psi4} package,\cite{Sherrill20_184108} and Mazziotti and coworkers have developed a code that is interfaced to Maple.\cite{Mazziotti20_3658} Third, the v2RDM literature tends to use language and notations that may seem foreign to ``traditional'' electronic structure theorists. This tutorial review seeks to address each of these issues, placing a particular emphasis on casting the v2RDM problem in a way that is digestible for researchers with a general background in electronic structure theory, while also providing Python code for implementing the v2RDM method and interfacing it with \textsc{Psi4}.

\section{Theory}

\label{SEC:THEORY}

This section outlines the theory underlying the variational optimization of the elements of the 2RDM under ensemble-state $N$-representability conditions. All quantities are represented in a basis of orthonormal spin orbitals, $\phi_{p_\sigma}$, where the labels $p$ and $\sigma \in \{\alpha, \beta\}$ represent the spatial and spin components of the orbital, respectively. 

\subsection{The electronic energy is a functional of the 2RDM}

Within the Born-Oppenheimer approximation, the non-relativistic Hamiltonian for a many-electron system has the form
\begin{equation}
    \label{EQN:H}
    \hat{H} = \sum_{pq} \sum_\sigma (T_{pq} + V_{pq})\hat{a}_{p_\sigma}^{\dagger}\hat{a}_{q_\sigma}+\frac{1}{2}\sum_{pqrs} \sum_{\sigma \tau}(pr|qs) \hat{a}_{p_\sigma}^{\dagger}\hat{a}_{q_\tau}^{\dagger}\hat{a}_{s_\tau}\hat{a}_{r_\sigma}
\end{equation}
Here, the symbols $T_{pq}$ and $V_{pq}$ represent integrals over the electron kinetic energy and electron-nucleus potential energy operators, respectively, $(pr|qs)$ denotes a two-electron repulsion integral in chemists' notation, and $\hat{a}_{p_\sigma}^{\dagger}$  ($\hat{a}_{p_\sigma}$) is a fermionic creation (annihilation) operator for spin orbital $\phi_{p_\sigma}$.

We now recognize that the expectation value of Eq.~\ref{EQN:H} is an exact functional of the one-electron reduced density matrix (1RDM), ${}^1{\bf D}$, and the 2RDM, ${}^2{\bf D}$. In second-quantized form, the elements of these RDMs are defined as
\begin{equation}
\label{EQN:D1}
    {}^1D^{p_\sigma}_{q_\sigma} = \langle \Psi | \hat{a}^\dagger_{p_\sigma}\hat{a}_{q_\sigma} |\Psi\rangle
\end{equation}
and
\begin{equation}
\label{EQN:D2}
    {}^2D^{p_\sigma q_\tau}_{r_\sigma s_\tau} = \langle \Psi | \hat{a}^\dagger_{p_\sigma}\hat{a}^\dagger_{q_\tau} \hat{a}_{ s_\tau} \hat{a}_{r_\sigma}|\Psi\rangle
\end{equation}
respectively. Given these definitions, it is easy to see that the electronic energy is expressible as
\begin{equation}
    \label{EQN:E}
    E = \sum_{pq} \sum_\sigma (T_{pq} + V_{pq}) {}^1D^{p_\sigma}_{q_\sigma}+\frac{1}{2}\sum_{pqrs} \sum_{\sigma \tau}(pr|qs) {}^2D^{p_\sigma q_\tau}_{r_\sigma s_\tau}
\end{equation}
As will be shown in section \ref{SEC:SYMMETRIES}, the elements of the 1RDM are expressible in terms of a partial trace over the 2RDM, so the energy can actually be thought of as simply a functional of the 2RDM. The question now becomes, is it possible to determine the 2RDM directly, via minimization of Eq.~\ref{EQN:E} with respect to variations in its elements? The answer is, of course, yes, provided that we can place enough restrictions on the 2RDM to guarantee that it comes from a single $N$-electron density matrix (pure-state $N$-representability) or an ensemble of such density matrices (ensemble $N$-representability). Sections \ref{SEC:SYMMETRIES} and \ref{SEC:POSITIVITY} detail a set of necessary, yet insufficient, ensemble $N$-representability conditions. Given that these conditions are not sufficient to guarantee the 2RDM is exactly ensemble-state $N$-representable (for systems with more than two electrons), the energy associated with a variationally obtained 2RDM will be a {\em lower}-bound to the exact energy ({\em i.e.}, the full configuration interaction [CI] energy obtained within the same one-electron basis set). 

\subsection{Encoding symmetries on the RDMs}
\label{SEC:SYMMETRIES}

The elements of the 1RDM and 2RDM should reflect the fundamental symmetries possessed by the state from which they derive. For example, the wave function should be antisymmetric with respect to the exchange of particle labels. This antisymmetry is reflected in the 2RDM, based on the commutation properties of the creation / annihilation operators that define it, {\em i.e.}
\begin{equation}
    {}^2D^{p_\sigma q_\tau}_{r_\sigma s_\tau} = -{}^2D^{q_\tau p_\sigma}_{r_\sigma s_\tau} = {}^2D^{p_\sigma q_\tau}_{s_\tau r_\sigma} = {}^2D^{q_\tau p_\sigma}_{s_\tau r_\sigma}
\end{equation}
Second, because the Hamiltonian in Eq.~\ref{EQN:H} is non-relativistic, the ground-state wave function $|\Psi\rangle$ could be an eigenfunction of the spin squared ($\hat{S}^2$) and $z$-projection of spin ($\hat{S}_z$) operators. In this case, the non-zero blocks of RDMs derived from such a state should reflect this spin symmetry, which is why we only considered the spin-conserving blocks of ${}^1{\bf D}$ and ${}^2{\bf D}$ in Eqs.~\ref{EQN:D1}-\ref{EQN:D2}. Such spin symmetry considerations correspond to only a subset of conditions that one could apply to the elements of the 1RDM and 2RDM based on the symmetries satisfied by $|\Psi\rangle$. 

Let us consider a state $|\Psi\rangle$ with fixed numbers of $\alpha$- and $\beta$-spin electrons and well-defined square of spin. Such a wave function satisfies the following equality constraints (among others)
\begin{align}
\label{EQN:N}
    \hat{N}_\sigma |\Psi\rangle &= N_\sigma |\Psi\rangle \\
    \label{EQN:N2}
    \hat{N}_\sigma \hat{N}_\tau|\Psi\rangle &= N_\sigma N_\tau|\Psi\rangle \\
    \label{EQN:S2}
    \hat{S}^2 |\Psi\rangle &= S(S+1) | \Psi\rangle
\end{align}
where $S$ is the spin angular momentum quantum number, 
\begin{align}
    \hat{N}_\sigma &= \sum_p \hat{a}^\dagger_{p_\sigma} \hat{a}_{p_\sigma} \\
    \hat{S}^2 &= \hat{S}_z + \hat{S}_z^2 + \hat{S}_{-}\hat{S}_{+}
\end{align}
and
\begin{align}
    \hat{S}_z &= \frac{1}{2}(\hat{N}_\alpha - \hat{N}_\beta) \\
    \hat{S}_{-} &= \sum_p \hat{a}^\dagger_{p_\beta} \hat{a}_{p_\alpha} \\
    \hat{S}_{+} &= \sum_p \hat{a}^\dagger_{p_\alpha} \hat{a}_{p_\beta}
\end{align}
By projecting Eqs.~\ref{EQN:N} and \ref{EQN:N2} onto the bra state $\langle \Psi|$ and bringing the creation / annihilation operators to normal order with respect to the true vacuum state, we obtain two sets of trace constraints on the spin blocks of the 1RDM and 2RDM, {\em i.e.}
\begin{align}
    \label{EQN:TR_D1}
    \sum_p {}^1D^{p_\sigma}_{p_\sigma} &= N_\sigma \\
    \label{EQN:TR_D2}
    \sum_{pq} {}^2D^{p_\sigma q_\tau}_{p_\sigma q_\tau} &= N_\sigma (N_\tau - \delta_{\sigma\tau})
\end{align}
Put together, these constraints guarantee that the variance in $\hat{N}_\alpha$, $\hat{N}_\beta$, and $\hat{S}_z$ are all zero when evaluated in terms of the 1RDM and 2RDM. Now, projecting Eq.~\ref{EQN:S2} on the bra state $\langle \Psi |$, bringing all operators to normal order, and replacing traces over the 1RDM and 2RDM with appropriate values (the right-hand sides of Eqs.~\ref{EQN:TR_D1} and \ref{EQN:TR_D2}), we obtain
\begin{equation}
    \sum_{pq} {}^2D^{p_\alpha q_\beta}_{q_\alpha p_\beta} = \frac{1}{2}(N_\alpha + N_\beta) + \frac{1}{4} (N_\alpha - N_\beta)^2 -S(S-1) 
\end{equation}
Additional spin symmetry constraints could be developed in terms of spin-adapted RDMs.\cite{Mazziotti05_052505, Ayers12_014110} Other expectation-value-type constraints can be defined if $|\Psi\rangle$ satisfies additional fundamental symmetries. For example, Refs.~\citenum{DePrince19_032509} and \citenum{DePrince22_5966} consider expectation value constraints involving the orbital angular momentum squared ($\hat{L}^2$), the $z$-projection of the orbital angular momentum ($\hat{L}_z$), the total angular momentum squared ($\hat{J}^2$), and the $z$-projection of the total angular momentum ($\hat{J}_z$). 

Additional constraints relating the elements of the 1RDM and the 2RDM can be derived from Eq.~\ref{EQN:N}. For example, consider the projection of Eq.~\ref{EQN:N} onto a different bra state, $\langle \Psi| \hat{a}^\dagger_{p_\sigma} \hat{a}_{q_\sigma}$
\begin{equation}
    \langle \Psi| \hat{a}^\dagger_{p_\tau} \hat{a}_{q_\tau} \hat{N}_\sigma | \Psi \rangle = N_\sigma \langle \Psi| \hat{a}^\dagger_{p_\tau} \hat{a}_{q_\tau} | \Psi \rangle
\end{equation}
which leads to the contraction constraint
\begin{equation}
\label{EQN:D2_TO_D1}
    \sum_r {}^2D^{p_\tau r_\sigma}_{q_\tau r_\sigma} = (N_\sigma - \delta_{\sigma \tau}) {}^1D^{p_\tau}_{q_\tau}
\end{equation}
Now, it is clear that the 1RDM in Eq.~\ref{EQN:E} could be replaced by a partial trace over the 2RDM such that the energy is expressible only in terms of the 2RDM.

\subsection{Additional ensemble $N$-representability conditions}
\label{SEC:POSITIVITY}

The constraints outlined in section \ref{SEC:SYMMETRIES} encode the particle number and spin symmetry of the state on the 1RDM and 2RDM. Physically meaninful RDMs are also positive semidefinite. The eigenvalues of the 1RDM and 2RDM are natural orbital and geminal occupation numbers, respectively, which by definition must be non-negative. The positivity requirement can also be appreciated from a simple mathematical perspective.\cite{Erdahl01_042113} 
Consider a set of operators, $\hat{C}^\dagger_I$, that can generate a basis of functions out of the state $|\Psi\rangle$ as
\begin{equation}
    |\chi_I\rangle = \hat{C}^\dagger_I | \Psi \rangle
\end{equation}
The overlap or Gram matrix associated with this basis, with elements $M_{IJ} = \langle \chi_I | \chi_J \rangle$, must be non-negative, by definition, which we denote as
\begin{equation}
    M \succeq 0
\end{equation}
Now, we can see that the non-negativity of the $\sigma$-spin block of the 1RDM is implied by the choice $\hat{C}^\dagger_I = \hat{a}_{q_\sigma}$, while that of the $\sigma\tau$-spin block of the 2RDM is implied by $\hat{C}^\dagger_I = \hat{a}_{s_\tau} \hat{a}_{r_\sigma}$. We are also free to consider other choices for $\hat{C}_I^\dagger$ that imply the non-negativity of the eigenvalues of additional RDMs. For example, $\hat{C}^\dagger_I = \hat{a}^\dagger_{q_\sigma}$ implies the non-negativity of the one-hole RDM, ${}^1{\bf Q}$, with elements
\begin{equation}
    \label{EQN:Q1}
    {}^1Q^{q_\sigma}_{p_\sigma} = \langle \Psi | \hat{a}_{q_\sigma} \hat{a}^\dagger_{p_\sigma} | \Psi \rangle
\end{equation}
We can use this knowledge to place additional constraints on the 1RDM that can be derived from the anticommutation relation
\begin{equation}
    \label{EQN:ANTICOMMUTATOR}
    \hat{a}^\dagger_{p_\sigma} \hat{a}_{q_\tau} +  \hat{a}_{q_\tau}\hat{a}^\dagger_{p_\sigma} = \delta_{pq}\delta_{\sigma \tau}
\end{equation}
Equations \ref{EQN:Q1} and \ref{EQN:ANTICOMMUTATOR} tell us that, not only should ${}^1{\bf D}$ {\color{black}be} positive semidefinite, but the matrix defined by the elements
\begin{equation}
    {}^1Q^{q_\sigma}_{p_\sigma} = \delta_{pq} - {}^1D^{p_\sigma}_{q_\sigma}
\end{equation}
should be as well. Put together, ${}^1{\bf Q} \succeq 0$ and ${}^1{\bf D} \succeq 0$ guarantee that the eigenvalues of the 1RDM will lie between zero and one. Such a 1RDM is ensemble-state $N$-representable.\cite{Coleman63_668} Before moving on, we note that the spin-block structure of ${}^1{\bf Q}$ is the same as that for ${}^1{\bf D}$.

Table \ref{TAB:POSITIVITY} details the unique choices for $\hat{C}^\dagger_I$ that generate {\color{black}positivity} conditions on up to three-body RDMs.
\begin{table*}[!htpb]
    \caption{Generators for {\color{black}positivity} conditions involving one-, two-, and three-body RDMs.}
    \label{TAB:POSITIVITY}
    \begin{center}
    \setlength\extrarowheight{3pt}
    \begin{tabular}{l l l}
    \hline\hline
    $\hat{C}^\dagger_I$ & RDM type & RDM definition \\
    \hline
    $\hat{a}_{q_\sigma}$           & one-electron & ${}^1D^{p_\sigma}_{q_\sigma} = \langle \Psi | \hat{a}^\dagger_{p_\sigma} \hat{a}_{q_\sigma} | \Psi \rangle$ \\
    $\hat{a}^\dagger_{q_\sigma}$   & one-hole     & ${}^1Q^{p_\sigma}_{q_\sigma} = \langle \Psi | \hat{a}_{p_\sigma} \hat{a}^\dagger_{q_\sigma} | \Psi \rangle$ \\
    $\hat{a}_{s_\tau}\hat{a}_{r_\sigma}$           & two-electron & ${}^2D^{p_\sigma q_\tau}_{r_\sigma s_\tau} = \langle \Psi | \hat{a}^\dagger_{p_\sigma} \hat{a}^\dagger_{q_\tau}\hat{a}_{s_\tau}\hat{a}_{r_\sigma} | \Psi \rangle$ \\
    $\hat{a}^\dagger_{s_\tau}\hat{a}^\dagger_{r_\sigma}$           & two-hole & ${}^2Q^{p_\sigma q_\tau}_{r_\sigma s_\tau} = \langle \Psi | \hat{a}_{p_\sigma} \hat{a}_{q_\tau}\hat{a}^\dagger_{s_\tau}\hat{a}^\dagger_{r_\sigma} | \Psi \rangle$ \\
    $\hat{a}^\dagger_{s_\lambda}\hat{a}_{r_\kappa}$           & electron-hole & ${}^2G^{p_\sigma q_\tau}_{r_\kappa s_\lambda} = \langle \Psi | \hat{a}^\dagger_{p_\sigma} \hat{a}_{q_\tau}\hat{a}^\dagger_{s_\lambda}\hat{a}_{r_\kappa} | \Psi \rangle$ \\
    $\hat{a}_{u_\kappa}  \hat{a}_{t_\tau}  \hat{a}_{s_\sigma}$ & three-electron & $^3D^{p_\sigma q_\tau r_\kappa}_{s_\sigma t_\tau u_\kappa} =
              \langle \Psi |
              \hat{a}^\dagger_{p_\sigma}  \hat{a}^\dagger_{q_\tau} \hat{a}_{r_\kappa}^{\dagger}
              \hat{a}_{u_\kappa}  \hat{a}_{t_\tau}  \hat{a}_{s_\sigma} 
              | \Psi \rangle$ \\
    $\hat{a}_{u_\nu}^{\dagger}  \hat{a}_{t_\mu}  \hat{a}_{s_\lambda}$  & electron-electron-hole & $^3E^{p_\sigma q_\tau r_\kappa}_{ s_\lambda t_\mu u_\nu} =
              \langle \Psi |
              \hat{a}^\dagger_{p_\sigma}  \hat{a}^\dagger_{q_\tau} \hat{a}_{r_\kappa}
              \hat{a}_{u_\nu}^{\dagger}  \hat{a}_{t_\mu}  \hat{a}_{s_\lambda} 
              | \Psi \rangle$ \\
    $\hat{a}^\dagger_{p_\sigma}  \hat{a}^\dagger_{q_\tau} \hat{a}_{r_\kappa}$ & electron-hole-hole & $^3F^{p_\sigma q_\tau r_\kappa}_{ s_\lambda t_\mu u_\nu} =
            \langle \Psi |
            \hat{a}_{u_\nu}^{\dagger}  \hat{a}_{t_\mu}  \hat{a}_{s_\lambda}
            \hat{a}^\dagger_{p_\sigma}  \hat{a}^\dagger_{q_\tau} \hat{a}_{r_\kappa}
            | \Psi \rangle$ \\
      $\hat{a}^\dagger_{u_\kappa}  \hat{a}^\dagger_{t_\tau}  \hat{a}^{\dagger}_{s_\sigma}$  & three-hole & $^3Q^{p_\sigma q_\tau r_\kappa}_{ s_\sigma t_\tau u_\kappa}=
              \langle \Psi |
              \hat{a}_{p_\sigma}  \hat{a}_{q_\tau} \hat{a}_{r_\kappa}
              \hat{a}^\dagger_{u_\kappa}  \hat{a}^\dagger_{t_\tau}  \hat{a}^{\dagger}_{s_\sigma} 
              | \Psi \rangle$ \\    
    \hline
 
\hline\hline
\end{tabular}
\end{center}
\end{table*}
In Table \ref{TAB:POSITIVITY}, we find two additional two-body RDMs: the two-hole RDM, ${}^2{\bf Q}$, which has the same spin-block structure as ${}^2{\bf D}$, and the electron-hole RDM, ${}^2{\bf G}$, whose spin-block structure is more complex. The non-zero spin blocks of ${}^2\mathbf{G}$ are those for which the number of $\alpha$-spin ($\beta$-spin) creation operators equals the number of $\alpha$-spin ($\beta$-spin) annihilation operators, which results in the spin-block structure
\begin{equation}
{}^2{\bf G} =
 \begin{pmatrix}
  {}^2{\bf G}^{\alpha\alpha}_{\alpha\alpha}  & {}^2{\bf G}^{\alpha\alpha}_{\beta\beta}   &  0  &  0 \\
  {}^2{\bf G}^{\beta\beta}_{\alpha\alpha}    & {}^2{\bf G}^{\beta\beta}_{\beta\beta}     &  0  &  0 \\
             0 &    0 &  {}^2{\bf G}^{\alpha\beta}_{\alpha\beta}  & 0 \\
             0 &    0 & 0 & {}^2{\bf G}^{\beta\alpha}_{\beta\alpha} \\
 \end{pmatrix}
\end{equation}
The combined non-negativity of ${}^2{\bf D}$, ${}^2{\bf Q}$, and ${}^2{\bf G}$, along with appropriate constraints enforcing the mapping between their elements implied by Eq.~\ref{EQN:ANTICOMMUTATOR}, constitute the PQG or DQG constraints of Garrod and Percus.\cite{Percus64_1756} With a good implementation, PQG-quality calculations can be carried out on large numbers of strongly correlated electrons. As an example, Refs.~\citenum{Mazziotti10_014104} and \citenum{DePrince19_6164} consider systems of up to 64 electrons correlated among 64 orbitals. As such, v2RDM with PQG conditions can be used as a solver for large-active-space CASSCF calculations, which can provide insight into the qualitative properties of large, strongly correlated systems, and quantitative predictions of energy differences such as spin state splittings can be quite accurate.\cite{DePrince16_2260} However, the accuracy of PQG-derived absolute energies can be poor,\cite{Mazziotti_11_052506} which motivates the consideration of higher-order conditions. 

Table \ref{TAB:POSITIVITY} introduces four three-body RDMs whose mutual non-negativity constitutes full three-positivity or 3POS.\cite{Mazziotti06_032501,DePrince21_174110} Note that, in 3POS calculations, one must also impose proper mappings between the spin blocks of the three-electron RDM (3RDM) and those of the 2RDM, which can be generated by projecting Eq.~\ref{EQN:N} onto the bra state $\langle \Psi | \hat{a}^\dagger_{p_\sigma}\hat{a}^\dagger_{q_\tau} \hat{a}_{s_\tau}\hat{a}_{r_\sigma}$. The floating-point and storage requirements for full 3POS calculations are high: $\mathcal{O}(n^9)$ and $\mathcal{O}(n^6)$, respectively, where $n$ is the number of active / correlated orbitals.  As such, it is useful to consider weaker but more manageable three-body constraints on the 2RDM by considering judiciously chosen linear combinations of the three-body RDMs,\cite{Erdahl78_697, Percus04_2095} which should also be positive semidefinite:
\begin{align}
    \label{EQN:T1}
    \mathbf{T1} = {}^3\mathbf{D} + {}^3\mathbf{Q}\\
    \label{EQN:T2}
    \mathbf{T2} = {}^3\mathbf{E} + {}^3\mathbf{F}
\end{align}
Importantly, the right-hand sides of Eqs.~\ref{EQN:T1} and \ref{EQN:T2} can be evaluated with knowledge of only the 1RDM and 2RDM, which removes the need to store the 3RDM or enforce any contraction constraints between it and the 2RDM. {\color{black}For example, ignoring the spin labels, the T2 matrix is
\begin{align}
    T2^{pqr}_{stu} &= -{}^2D^{pq}_{ts} \delta_{ru} - {}^2D^{qu}_{tr} \delta_{ps} + {}^2D^{qu}_{sr} \delta_{pt} + {}^2D^{pu}_{tr} \delta_{qs} - {}^2D^{pu}_{sr} \delta_{qt} \nonumber \\ 
    &+ {}^1D^u_r \delta_{ps}\delta_{qt} - {}^1D^u_r \delta_{pt} \delta_{qs}
\end{align}
}In practice, the T2 constraint is much stronger than the T1 constraint, so the latter can be ignored without any significant energetic consequences. Like ${}^2{\bf G}$, the {\bf T2} matrix has a much more complex spin-block structure than the 3RDM (or ${}^3{\bf Q}$ or the {\bf T1} matrix), which reflects the spin-block structure of the  ${}^3\mathbf{E}$ and ${}^3\mathbf{F}$ RDMs. The non-zero blocks of the {\bf T2} matrix are those for which the number of $\alpha$-spin ($\beta$-spin) creation operators equals the number of $\alpha$-spin ($\beta$-spin) annihilation operators, or
\begin{equation}
{\bf T2} =
 \begin{pmatrix}
  {\bf T2}^{\alpha\alpha\alpha}_{\alpha\alpha\alpha}  & {\bf T2}_{\alpha\alpha\alpha}^{\alpha\beta\beta} &  0  &  0  & 0 & 0 \\
  {\bf T2}_{\alpha\beta\beta}^{\alpha\alpha\alpha}    & {\bf T2}^{\alpha\beta\beta}_{\alpha\beta\beta}   &  0  &  0  & 0 & 0 \\
  0 & 0 & {\bf T2}^{\beta\alpha\alpha}_{\beta\alpha\alpha} & {\bf T2}_{\beta\alpha\alpha}^{\beta\beta\beta} & 0 & 0 \\
  0 & 0 & {\bf T2}_{\beta\beta\beta}^{\beta\alpha\alpha}   & {\bf T2}^{\beta\beta\beta}_{\beta\beta\beta}   & 0 & 0 \\
  0 & 0 & 0 & 0 & {\bf T2}^{\alpha\alpha\beta}_{\alpha\alpha\beta}   & 0 \\
  0 & 0 & 0 & 0 & 0 & {\bf T2}^{\beta\beta\alpha}_{\beta\beta\alpha}  \\
 \end{pmatrix}
\end{equation}

In principle, higher-order conditions involving four-particle RDMs, etc., could be developed, and, as we further constrain the 2RDM, the v2RDM energy approaches that from full CI from below, with the two approaches being equivalent in the limit that  the non-negativity of the eigenvalues of the full $N$-particle RDM is maintained. Such calculations would be intractable in general, though, so we typically truncate the hierarchy at the two-particle (PQG) or three-particle (3POS or T1/T2) level. We also note that the full CI limit can also be reached by considering partial high-order conditions similar to T1/T2, where linear combinations of high-order RDMs are taken such that the resulting constraints apply to quantities that are expressible only in terms of the 2RDM.\cite{Mazziotti12_263002, Mazziotti12_062507, Mazziotti23_153001}

\subsection{The v2RDM semidefinite program}

Minimizing Eq.~\ref{EQN:E} with respect to variations in the elements of the 2RDM involves a large number of linear constraints, as well as enforcing the non-negativity of the 2RDM and the other RDMs described in section \ref{SEC:POSITIVITY}. The optimization can be cast as a semidefinite programming (SDP) problem, and substantial effort has gone into the development of SDP solvers that can optimize 2RDMs for many-body systems. Early implementations\cite{Fujisawa01_8282,Mazziotti02_062511,Mazziotti06_032501} demonstrated the potential power of v2RDM theory, but the computational costs of the primal-dual interior-point algorithms employed at the time were too great for the approach to be competitive with wave-function-based methods. When enforcing PQG $N$-representability conditions, the interior-point algorithms require $\mathcal{O}(n^{16})$ floating-point operations and $\mathcal{O}(n^8)$ storage, where $n$ is the number of orbitals correlated. A series of breakthroughs have substantially reduced these costs. For example, Mazziotti has presented both a matrix-factorization-based\cite{Mazziotti04_213001} algorithm (RRSDP) and a boundary-point\cite{Wiegele06_277,Wiegele09_336,Mazziotti11_083001} algorithm (BPSDP) that reduce these floating-point and memory requirements to $\mathcal{O}(n^6)$ and $\mathcal{O}(n^4)$, respectively, with the latter algorithm displaying a lower prefactor. Canc\`{e}s, Stoltz, and Lewin\cite{Lewin06_064101} have also developed a dual-cone optimization algorithm that achieves similar scalings. Mazziotti has developed his own low-rank\cite{Mazziotti16_153001} dual-cone optimization of the 2RDM\cite{Mazziotti20_052819} based on RRSDP that allows for the application of partial three-fermion $N$-representability conditions ({\em e.g} the T2 constraint\cite{Erdahl78_697,Percus04_2095}) at a similar $\mathcal{O}(n^6)$ cost{\color{black}; Ref.~\citenum{Mazziotti16_153001} treats a 30-orbital active space with the reduced-rank T2 constraint.} In this tutorial review, we focus on the RRSDP and BPSDP algorithms, which we have implemented in the library \texttt{libSDP}.\cite{libsdp} 

We begin by reviewing the primal and dual formulations of the SDP problem. The primal formulation is
\begin{eqnarray}
\label{EQN:PRIMAL}
\min_{\bf x} &{\bf c}^T{\bf x}\\\notag
\text{such that }& {\bf A}{\bf x} &= {\bf b}\\\notag
\text{and }& M({\bf x}) &\succeq 0,
\end{eqnarray}
Here, ${\bf x}$ is the primal solution vector, ${\bf c}$ is a vector containing the the 1- and 2-electron integrals that make up the Hamiltonian, and ${\bf A}$ and ${\bf b}$ represent the constraint matrix and vector, respectively, which, together, encode the $N$-representability conditions applied to the 2RDM. Each row of the matrix ${\bf A}$ represents a constraint, and the action of ${\bf A}$ on ${\bf x}$ plucks out the appropriate RDM elements contributing to that constraint. For example, for the trace constraints given in Eq.~\ref{EQN:N}, the relevant rows of {\bf A} consists of 1's in the positions corresponding to the diagonal elements of ${}^1{\bf D}_\sigma$ and 0's elsewhere, and the relevant elements of ${\bf b}$ are $N_\sigma$.  $M({\bf x})$ provides a mapping between the primal solution vector and the RDMs as
\begin{equation}
\label{EQN:MAPPING}
M({\bf x})=\left(\begin{array}{cccccc}{}^1{\bf D}&0&0&0&0&0\\
                           0&{}^1{\bf Q}&0&0&0&0\\
                           0&0&{}^2{\bf D}&0&0&0\\
                           0&0&0&{}^2{\bf Q}&0&0\\
                           0&0&0&0&{}^2{\bf G}&0\\
                           0&0&0&0&0&...\\
                           \end{array}\right) \succeq 0.
\end{equation}
We note that the elements of ${\bf c}$ should be arranged such that the inner product ${\bf
x}^{T}{\bf c}$ yields the electronic energy given by Eq.~\ref{EQN:E}.

The dual formulation of the SDP problem takes the form
\begin{eqnarray}
\label{EQN:DUAL}
\max_{\bf y} & {\bf b}^T{\bf y}\\\notag
\text{such  that } &{\bf z}&  = {\bf c} - {\bf A}^T {\bf y} \\\notag
\text{and }        &M({\bf z})& \succeq 0
\end{eqnarray}
where the pair of vectors (${\bf y}$,${\bf z}$) forms the dual solution, and $M({\bf z})$ maps ${\bf z}$ onto a set of matrices with the same structure as the RDMs found in Eq.~\ref{EQN:MAPPING}. In addition to the primal and dual constraints given in Eqs.~\ref{EQN:PRIMAL} and \ref{EQN:DUAL}, the simultaneous optimality of ${\bf x}$ and (${\bf y}$, ${\bf z}$) also requires  the inner product ${\bf x}^T{\bf z}$ to vanish.

Both the primal and dual forms of the SDP represent constrained optimizations that can be  solved using augmented Lagrangian techniques. When doing so, the primal and dual solutions are complements of one another in the sense that the dual variable ${\bf y}$ acts as the Lagrange multiplier within the primal problem, while the primal variable ${\bf x}$ acts the Lagrange multiplier within the dual problem. As such, we define the augmented Lagrangian for the primal problem problem as
\begin{equation}
\label{EQN:PRIMAL_LAGRANGIAN}
\mathcal{L}_{\rm P} = {\bf c}^T {\bf x} - {\bf y}^T({\bf A} {\bf x}-{\bf b}) + \frac{1}{2\mu} ||{\bf A} {\bf x}-{\bf b}||^2    
\end{equation}
where the elements of ${\bf y}$ are the Lagrange multipliers, and $\mu$ is a non-negative penalty parameter. The augmented Lagrangian for the dual problem is
\begin{equation}
\label{EQN:DUAL_LAGRANGIAN}
\mathcal{L}_{\rm D} = {\bf b}^T {\bf y} - {\bf x}^T({\bf A}^T {\bf y}-{\bf c}+{\bf z}) + \frac{1}{2\mu} ||{{\bf A}^T {\bf y}-{\bf c}+{\bf z}}||^2
\end{equation}
where $\mu$ is again a non-negative penalty parameter, and ${\bf x}$ contains the Lagrange multipliers. The RRSDP algorithm can be used to minimize $\mathcal{L}_{\rm P}$ with respect to ${\bf x}$, and the BPSDP algorithm can be used to maximize $\mathcal{L}_{\rm D}$ with respect to ${\bf y}$. In either case, the algorithms should be formulated so as to avoid the explicit storage of the full matrix ${\bf A}$; rather, one should store ${\bf A}$ in a sparse-matrix representation or, alternatively, directly implement the action of ${\bf A}$ or ${\bf A}^T$ on vectors of the dimension of the primal solution vector (the number of RDM elements) or the dual solution vector (the number of constraints), respectively.

\subsubsection{RRSDP} 

In the RRSDP algorithm, the non-negativity of the primal solution, $M({\bf x})$, is enforced through a matrix factorization
\begin{equation}
\label{EQN:RR}
M({\bf x}) = {\bf R}{\bf R}^T
\end{equation}
{\color{black}
With this choice, the dot product that is the primal energy can be expressed as
\begin{equation}
{\bf c}^T {\bf x} = \text{Tr}[M({\bf c}) M({\bf x})] = \text{Tr}[M({\bf c}) {\bf R}{\bf R}^T]
\end{equation}
and we rewrite the primal problem in terms of the auxiliary variable, ${\bf R}$
\begin{eqnarray}
\label{EQN:PRIMAL_R}
\min_{\bf R} & \text{Tr}[M({\bf c})M({\bf R}{\bf R}^T)]\\\notag
\text{such that }& {\bf A}{\bf x} = {\bf b}\\\notag
\text{and }& M({\bf x}) = {\bf R}{\bf R}^T,
\end{eqnarray}
}For the minimization itself, we use the limited-memory BFGS procedure (\texttt{libSDP} uses the in the implementation in the \texttt{libLBFGS} library\cite{liblbfgs}), which requires the gradient of the objective function with respect to ${\bf R}$. After some straightforward but tedious manipulations, this gradient is given by
{\color{black}\begin{equation}
\label{EQN:dLdR}
    \frac{\partial \mathcal{L}_{\rm P}}{\partial {\bf R}} = 2 {\bf R}^T M({\bf c} + {\bf A}^T[\frac{1}{\mu}({\bf A}{\bf x} - {\bf b})- {\bf y}])
\end{equation}
}Now, we can see that evaluating the objective function (Eq.~\ref{EQN:PRIMAL_LAGRANGIAN}) and its gradient (Eq.~\ref{EQN:dLdR}) requires evaluating (i) several dot products between vectors of the dimension of the primal or dual solution vectors, (ii) the action of ${\bf A}$ and ${\bf A}^T$ on vectors of the dimension of the primal and dual solution vectors, respectively, and (iii) the two tensor contractions in Eqs.~\ref{EQN:RR} and \ref{EQN:dLdR}. When enforcing the PQG conditions, the formal scaling of the first two sets of operations is $\mathcal{O}(n^4)$, and that of the tensor contractions is $\mathcal{O}(n^6)$. When enforcing three-particle conditions ({\em e.g.}, {T2}), these costs grow to $\mathcal{O}(n^6)$ and $\mathcal{O}(n^9)$, respectively.  In either case, the tensor contractions dominate the computational cost of the RRSDP algorithm.

Given an initial random guess for ${\bf R}$ and an initial non-negative value for $\mu$, the overall RRSDP procedure is as follows
\begin{enumerate}
    \item Minimize Eq.~\ref{EQN:PRIMAL_LAGRANGIAN} for fixed $\mu$ and ${\bf y}$ with respect to variations in ${\bf x}$.
    \item Update ${\bf y}$ and $\mu$ according to
    \begin{align}
        {\bf y} &= {\bf y} - \frac{1}{\mu}({\bf A}{\bf x} - {\bf b}) \\
        \mu &= 0.1 \times \mu
    \end{align}
    \item Repeat steps 1 and 2 until the primal error, $||{\bf A}{\bf x} - {\bf b}||$, and the change in the primal energy, ${\bf c}^T {\bf x}$, fall below some predefined thresholds.

\end{enumerate}

\subsubsection{BPSDP} 

The BPSDP algorith, originally developed by Weigele and coworkers,\cite{Wiegele06_277,Wiegele09_336} was adapted by Mazziotti for quantum chemical applications in 2012.\cite{Mazziotti11_083001} We follow Mazziotti's formulation of the algorithm, with only minor modifications. 
The primal and dual solutions, ${\bf x}$ and (${\bf y}$, ${\bf z}$), are determined through an iterative two-step procedure. Given initial $\mu > 0$, $M({\bf x}) \succeq 0$, and $M({\bf z}) \succeq 0$, Eq.~\ref{EQN:DUAL_LAGRANGIAN} should be maximized with respect to {\bf y}. The stationary condition $\partial\mathcal{L}_{\rm D} / \partial{\bf y} = 0$ leads to
\begin{equation}
\label{EQN:CG}
{\bf AA}^T{\bf y} = {\bf A}({\bf c}-{\bf z}) + \mu({\bf b}-{\bf A}{\bf x}),
\end{equation}
which can be solved via the conjugate gradient (CG) approach. The cost of the CG procedure is dominated by the repeated evaluation of the left-hand size of Eq.~\ref{EQN:CG}, which involves the evaluation of ${\bf d} = {\bf A}^T{\bf y}$, followed by ${\bf A}{\bf d}$. As in RRSDP, such operations carry $\mathcal{O}(n^4)$ or $\mathcal{O}(n^6)$ cost when enforcing the two- or three-body $N$-representability conditions, respectively. The actual CG algorithm itself carries comparable cost. Given ${\bf y}$, updated estimates for ${\bf x}$ and ${\bf z}$ are then obtained from the eigendecomposition of ${\bf U} = M(\mu {\bf x} + {\bf A}^T{\bf y} - {\bf c})$, {\em i.e.},
\begin{align}
    \label{EQN:UPDATE_XZ}
    {\bf x} &= \frac{1}{\mu} {\bf U}_+ \\
    {\bf z} &= - {\bf U}_-
\end{align}
where ${\bf U}_{\pm}$ represents the positive/negative components of the eigenspectrum of ${\bf U}$. Given this update, the conditions $M({\bf x}) \succeq 0$, $M({\bf z}) \succeq 0$, and ${\bf x}^T {\bf z} = 0$ are all satisfied. The formal scaling of the eigendecomposition is $\mathcal{O}(n^6)$ or $\mathcal{O}(n^9)$ when enforcing two- or three-body conditions, respectively. While the worst-scaling steps of the RRSDP and BPSDP are comparable, the prefactor for the BPSDP algorithm is significantly lower, which is why it is the preferred algorithm for large-scale v2RDM calculations. Moreover, in BPSDP, the actual walltime for PQG calculations tends to be dominated by the CG step, despite the scaling of that step being lower than that of the eigendecomposition. However, this behavior changes when the algorithm is implemented on an accelerator such as a graphical processing unit. In this case, the CG step is significantly more efficient on the GPU than on the CPU, so the eigendecomposition step ends up dominating the cost of the calculation for large systems.\cite{DePrince19_6164} The reason the CG step is amenable to accelaration on the GPU is that most of the work in this step takes the form of level-1 BLAS or other memory operations, which benefits from the greater memory bandwidth on the GPU, relative to a comparable quality CPU.

\section{The \texttt{libSDP} Library}

\label{SEC:LIBSDP}

We have implemented the RRSDP and BPSDP algorithms in an open-source library, \texttt{libSDP}, which is available on GitHub.\cite{libsdp} \texttt{libSDP} was originally developed as a C++ library to which one could represent any SDP problem by supplying (i) the vector ${\bf c}$ that defines the problem and (ii) the vector ${\bf b}$ encoding the constraint values, (iii) a list of block dimensions for each block of the RDMs that make up $M({\bf x})$, and (iv) suitable callback functions for the evaluation of the action of ${\bf A}$ and ${\bf A}^T$ on primal- or dual-sized vectors. This C++ library serves as the SDP solver for the v2RDM implementation in \texttt{Hilbert},\cite{hilbert} which is a plugin to the \textsc{Psi4}\cite{Sherrill20_184108} package. More recently, we have developed a Python interface to \texttt{libsdp} that allows one to define the SDP problem directly in Python. In this section, we develop a working minimal example for a two-electron system, H$_2$, in its lowest-energy singlet state, which only requires us to consider the $\alpha\beta$-block of the 2RDM to reach the full CI limit. More complete examples that implement the full PQG conditions with interfaces to \textsc{Psi4} and Pyscf can be found on GitHub.\cite{libsdp}

\begin{figure}[!htpb]
\caption{Python code snippet that carries out a Hartree-Fock calcualtion using the \textsc{Psi4} package and computes the one- and two-electron integrals in the MO basis.}
\label{FIG:SETUP_PSI4}
\begin{python}

import psi4
import numpy as np
# set molecule
mol = psi4.geometry("""
  0 1
  H 
  H 1 1.0
  symmetry c1
""")

# set options
psi4_options_dict = {
  'basis': 'sto-3g',
  'scf_type': 'pk',
}
psi4.set_options(psi4_options_dict)

# compute HF energy and wave function
scf_e, wfn = psi4.energy('SCF', return_wfn=True)

# molecular orbitals 
C = wfn.Ca()

# use Psi4's MintsHelper to generate integrals
mints = psi4.core.MintsHelper(wfn.basisset())
T = np.asarray(mints.ao_kinetic())
V = np.asarray(mints.ao_potential())
oei = np.einsum('uj,vi,uv', C, C, T + V)
tei = np.asarray(mints.mo_eri(C, C, C, C))

# number of orbitals and alpha/beta electrons
n = wfn.nmo()
nalpha = wfn.nalpha()
nbeta = wfn.nbeta()
\end{python}
\end{figure}

First, we can use \textsc{Psi4} (or Pyscf, etc.) to obtain the one- and two-electron integrals required to define the v2RDM problem (the vector ${\bf c}$). 
As stated above, we consider molecular hydrogen (H$_2$) in its lowest-energy singlet state, represented by a minimal (STO-3G) basis set. For this system, the full CI is reached by considering the non-negativity of the $\alpha\beta$-block of the 2RDM, in which case, the mapping, $M({\bf x})$, looks like
\begin{equation}
\label{EQN:MAPPING_H2}
M({\bf x})=\left(\begin{array}{ccc}{}^1{\bf D}_\alpha\\
                           0&{}^1{\bf D}_\beta\\
                           0&0&{}^2{\bf D}_{\alpha\beta}\\
                           \end{array}\right) \succeq 0.
\end{equation}
Ignoring point-group symmetry, the primal solution vector has three blocks of size, $n \times n$, $n \times n$, and $n^2 \times n^2$.
Figure \ref{FIG:SETUP_PSI4} uses the Python interface to \textsc{Psi4} to run a Hartree-Fock calculation on H$_2$ and to transform the one-electron integrals (\texttt{oei}) and two-electron integrals (\texttt{tei}) to the molecular-orbital (MO) basis.

Given \texttt{oei} and \texttt{tei}, we can begin to develop the required inputs for the SDP solver. In the Python interface to \texttt{libsdp}, the one- and two-electron integrals (${\bf c}$) and the constraints (the rows of ${\bf A}$, ${\bf A}_i$) are represented using \texttt{sdp\_matrix} objects, which have four attributes: the block number identifying a particular RDM block (unit offset), row and column numbers identifying the relevant elements of the RDM within the block (unit offset), and a value.  Given the simple structure for our model problem (Eq.~\ref{EQN:MAPPING_H2}), we can define the primal energy as an \texttt{sdp\_matrix} object using the code snippet in Fig.~\ref{FIG:SET_INTEGRALS}.
\begin{figure}[!htpb]
\caption{Python code snippet that represents the vector ${\bf c}$ as an \texttt{sdp\_matrix} object.}
\label{FIG:SET_INTEGRALS}
\begin{python}
# import libsdp library
import libsdp

c = libsdp.sdp_matrix()

# alpha-spin one-electron integrals (block 1)
for p in range (0, n):
  for q in range (0, n):
    c.block_number.append(1)
    c.row.append(p+1)
    c.column.append(q+1)
    c.value.append(oei[p][q])

# beta-spin one-electron integrals (block 2)
for p in range (0, n):
  for q in range (0, n):
    c.block_number.append(2)
    c.row.append(p+1)
    c.column.append(q+1)
    c.value.append(oei[p][q])

# two-electron integrals (block 3)
for p in range (0, n):
  for q in range (0, n):
    pq = p * n + q
    for r in range (0, n):
      for s in range (0, n):
        rs = r * n + s
        c.block_number.append(3)  
        c.row.append(pq+1)
        c.column.append(rs+1)
        c.value.append(tei[p][r][q][s]) 
F = [c]
\end{python}
\end{figure}
Note that the \texttt{sdp\_matrix} representation of ${\bf c}$ is given as the first element of an array of \texttt{sdp\_matrix} objects, \texttt{F}, which will also contain \texttt{sdp\_matrix} objects corresponding to the constraints.

Once the \texttt{sdp\_matrix} representation of ${\bf c}$ has been constructed, we can define each of the linear constraints that apply to the blocks of $M({\bf x})$ in a similar way. For example, the trace constraint on the $\alpha\beta$-block of the 2RDM (Eq.~\ref{EQN:TR_D2}) is represented in the Python code in Fig.~\ref{FIG:SET_TR_D2}.
\begin{figure}[!htpb]
\caption{Python code snippet that represents the trace constraint on the 2RDM (Eq.~\ref{EQN:TR_D2}).}
\label{FIG:SET_TR_D2}
\begin{python}
# D2ab trace constraint
Ai = libsdp.sdp_matrix()
for p in range (0, n):
  for q in range (0, n):
    pq = p * n + q
    Ai.block_number.append(3)  
    Ai.row.append(pq+1)
    Ai.column.append(pq+1)
    Ai.value.append(1.0) 
F.append(Ai)
b = [nalpha * nbeta]
\end{python}
\end{figure}
Here, \texttt{nalpha} and \texttt{nbeta} represent the number of $\alpha$- and $\beta$-spin electrons, respectively. As linear constraints are developed, they are appended to the end of the list of \texttt{sdp\_matrix} objects contained in \texttt{F}, and the corresponding values of ${\bf b}$ should be collected in a list as well. Figure \ref{FIG:SET_D2_TO_D1} implements the ${}^2{\bf D} \to {}^1{\bf D}$ contraction constraints (Eq.~\ref{EQN:D2_TO_D1}). As we can see, a separate constraint is defined for the mapping of the 2RDM to each element of the $\alpha$- and $\beta$-blocks of the 1RDM. 
\begin{figure}[!htpb]
\caption{Python code snippet that represents the contraction of the 2RDM to the 1RDM (Eq.~\ref{EQN:D2_TO_D1}).}
\label{FIG:SET_D2_TO_D1}
\begin{python}
# D2ab -> D1a contraction constraint
for p in range (0, n):
  for q in range (0, n):
    Ai = libsdp.sdp_matrix()
    for r in range (0, n):
      pr = p * n + r
      qr = q * n + r
        
      # D2ab
      Ai.block_number.append(3)  
      Ai.row.append(pr+1)
      Ai.column.append(qr+1)
      Ai.value.append(1.0) 
        
    # D1a
    Ai.block_number.append(1)  
    Ai.row.append(p+1)
    Ai.column.append(q+1)
    Ai.value.append(-nbeta) 
        
    F.append(Ai)
    b.append(0.0)
    
# D2ab -> D1b contraction constraint
for p in range (0, n):
  for q in range (0, n):
    Ai = libsdp.sdp_matrix()
    for r in range (0, n):
      rp = r * n + p
      rq = r * n + q
        
      # D2ab
      Ai.block_number.append(3)  
      Ai.row.append(rp+1)
      Ai.column.append(rq+1)
      Ai.value.append(1.0) 
        
    # D1b
    Ai.block_number.append(2)  
    Ai.row.append(p+1)
    Ai.column.append(q+1)
    Ai.value.append(-nalpha) 
        
    F.append(Ai)
    b.append(0.0)
\end{python}
\end{figure}

Given the list of \texttt{sdp\_matrix} objects, \texttt{F}, and constraint values, \texttt{b}, we call the SDP solver using the code snippet given in Fig.~\ref{FIG:CALL_LIBSDP_SOLVER}. Various options (the solver type, convergence criteria, etc.) can be controlled with the \texttt{sdp\_options} object; a complete set of options and their default settings can be found on GitHub.\cite{libsdp} Aside from \texttt{F}, \texttt{b}, and the options object, the only other information required by the solver is a list of dimensions for each block of the primal solution, $M({\bf x})$.  When the solver finishes, it returns the primal solution vector, ${\bf x}$. Figure \ref{FIG:CALL_LIBSDP_SOLVER} also shows how the dual solution $({\bf y}, {\bf z})$ can be obtained (assuming they are available, {\em i.e.}, that the solver is BPSDP) and how one can calculate the primal and dual energies (${\bf c}^T{\bf x}$ and ${\bf b}^T{\bf y}$, respectively) and the primal and dual errors ($||{\bf Ax}-{\bf b}||$ and $||{\bf c} - {\bf z} - {\bf A}^T{\bf y}||$, respectively).
\begin{figure}[!htpb]
\caption{Python code snippet that calls the \texttt{libsdp} SDP solver and evaluates the primal and dual energies and errors. }
\label{FIG:CALL_LIBSDP_SOLVER}
\begin{python}
# grab options object
options = libsdp.sdp_options()

# optional: change default options
...

# grab sdp solver object
sdp = libsdp.sdp_solver(options)

# define dimensions for each block of M(x)
dimensions = [n, n, n * n]

# solve the SDP problem
x = sdp.solve(b, F, dimensions, options.maxiter)

# get dual solution vectors (assuming BPSDP)
z = np.array(sdp.get_z())
y = np.array(sdp.get_y())

# get c (which is just the first element of F)
c = np.array(sdp.get_c())

# evaluate primal and dual energies
primal_energy = np.dot(c, x)
dual_energy = np.dot(b, y)

# evaluate primal and dual errors

# action of A^T on y
ATy = np.array(sdp.get_ATu(y))

# action of A on x 
Ax = np.array(sdp.get_Au(x))

dual_error = np.linalg.norm(c - z - ATy)
primal_error = np.linalg.norm(Ax - b)

\end{python}
\end{figure}

\section{Numerical Applications}

\label{SEC:RESULTS}

\begin{figure*}[!htpb]
    \begin{center}
            \caption{The (a) potential energy curves for the dissociation of
            molecular nitrogen using CI- and v2RDM-CASSCF,
            and the (b) error in the v2RDM-CASSCF energy, relative to
            CI-CASSCF. Reprinted with permission from {\em J. Chem. Theory Comput.} {\bf 12}, 2260-2271 (2016).  Copyright 2016 American Chemical Society.}
        \label{FIG:N2_PEC}
        \includegraphics[scale=0.5]{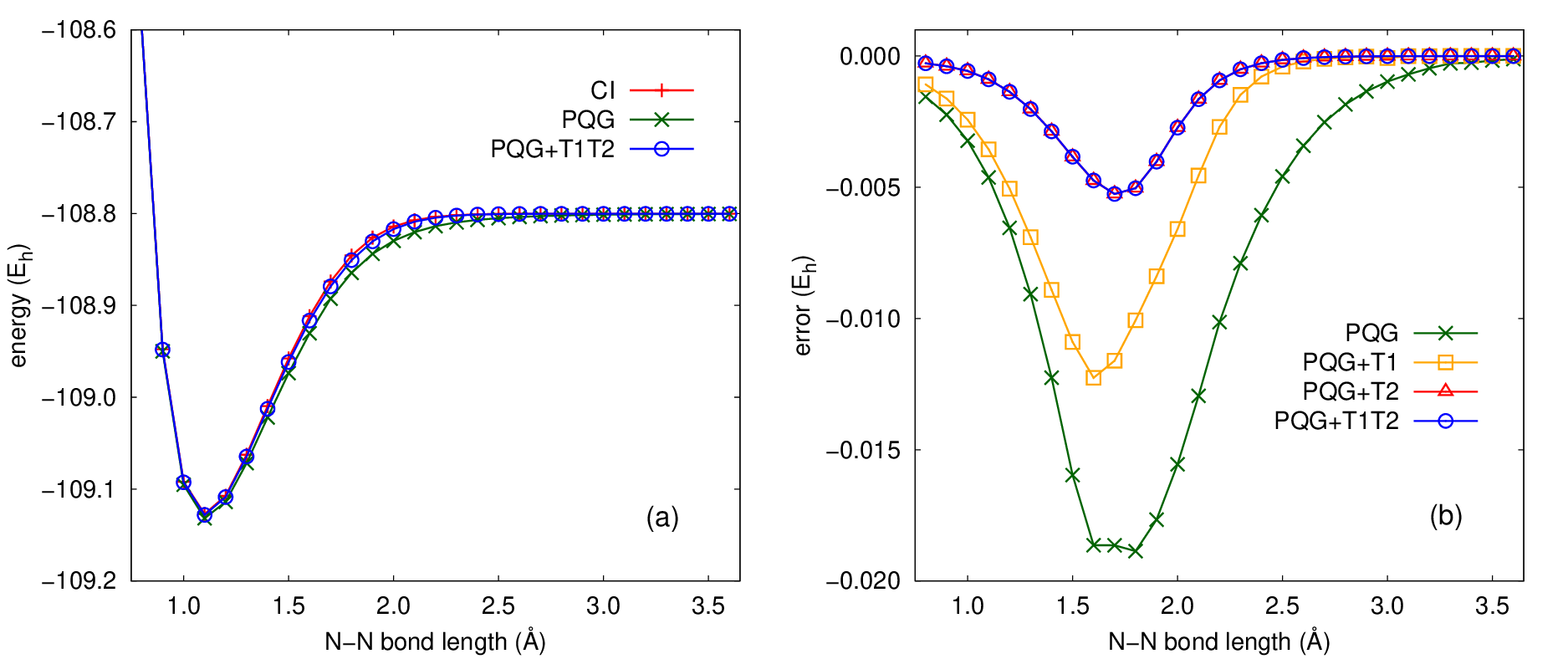}
    \end{center}
\end{figure*}

In this section, we discuss some of the numerical properties of the v2RDM approach. First, we consider a typical application of v2RDM theory, as a solver for the electronic structure of an active space within approximate CASSCF calculations. In this case, the active space energy is given by
\begin{align}
\label{EQN:ACTIVE_SPACE_ENERGY}
E_{\rm act} &= \frac{1}{2}\sum_{tuvw} (tv|uw) \sum_{\sigma\tau} {}^2D^{t_\sigma u_\tau}_{v_\sigma w_\tau} \nonumber \\
 &+ \sum_{tu} \bigg ( T_{tu} + V_{tu} + \sum_i [ 2 (tu|ii) - (ti|ui) ] \bigg ) \sum_{\sigma} {}^1D^{t_\sigma}_{u_\sigma}
\end{align}
where $t$, $u$, $v$, and $w$ represent active orbital labels, and $i$ is a label for an inactive occupied orbital. As mentioned earlier, the utility of the v2RDM approach is that it can be applied to large active spaces. For example, the implementation in Ref.~\citenum{DePrince19_6164} was applied to systems as large as 64 electrons distributed among 64 orbitals [a (64e, 64o) active space].

Figure \ref{FIG:N2_PEC} shows potential energy curves for the dissociation of molecular nitrogen (N$_2$), described by a cc-pVQZ basis set, as computed via CI- and v2RDM-driven CASSCF calculations. These data are taken from Ref.~\citenum{DePrince16_2260}. The calculations therein employed the
$D_{2h}$ point group and a (6e, 6o) active space, the specific definition of which can be found in that work. 
For this system, we see that v2RDM performs well in the dissociation limit, for all subsets of $N$-representability conditions considered. The largest-magnitude energy error observed is almost 20 mE$_{\rm h}$ at an N--N distance of 1.7~\AA, when using the PQG conditions. The T1 condition reduces this error to roughly 12 mE$_{\rm h}$, while the T2 condition displays a maximum error of only 5 mE$_{\rm h}$. As can be seen, the T1 condition is weak compared to T2, and application of T1 and T2 together does not improve over the application of T2 alone. As a result, many calculations consider only the application of the T2 condition. The accuracy of v2RDM for this case is typical for many small molecules containing second-row atoms. PQG conditions provide qualitative accuracy, in general, and more quantitative accuracy requires the application of partial or full three-particle conditions. 
Indeed, the correlation energy from PQG-quality calculations is often overestimated by as much as 20\%, even in small systems near their equilibrium geometries,\cite{Mazziotti_11_052506} while energies derived from PQG+T2 or full 3POS calculations can be accurate to 1 mE$_{\rm h}$ or less.\cite{Mazziotti06_032501,DePrince21_174110} As discussed below, though, in some challenging systems,\cite{Evangelista20_104108} even the application of full 3POS results in energy errors exceeding 20 mE${\rm h}$.\cite{DePrince21_174110}

\begin{figure*}[!htpb]
    \begin{center}
            \caption{Natural orbital occupation numbers for the singlet ground states of the (a)k-acene, (b) 3,k-circumacene, and (c) 3,k-periacene series. Adapted with permission from {\em J. Chem. Theory Comput.} {\bf 15}, 6164-6178-2271 (2019).  Copyright 2019 American Chemical Society.}
        \label{FIG:ACENES}
        \includegraphics[scale=0.35]{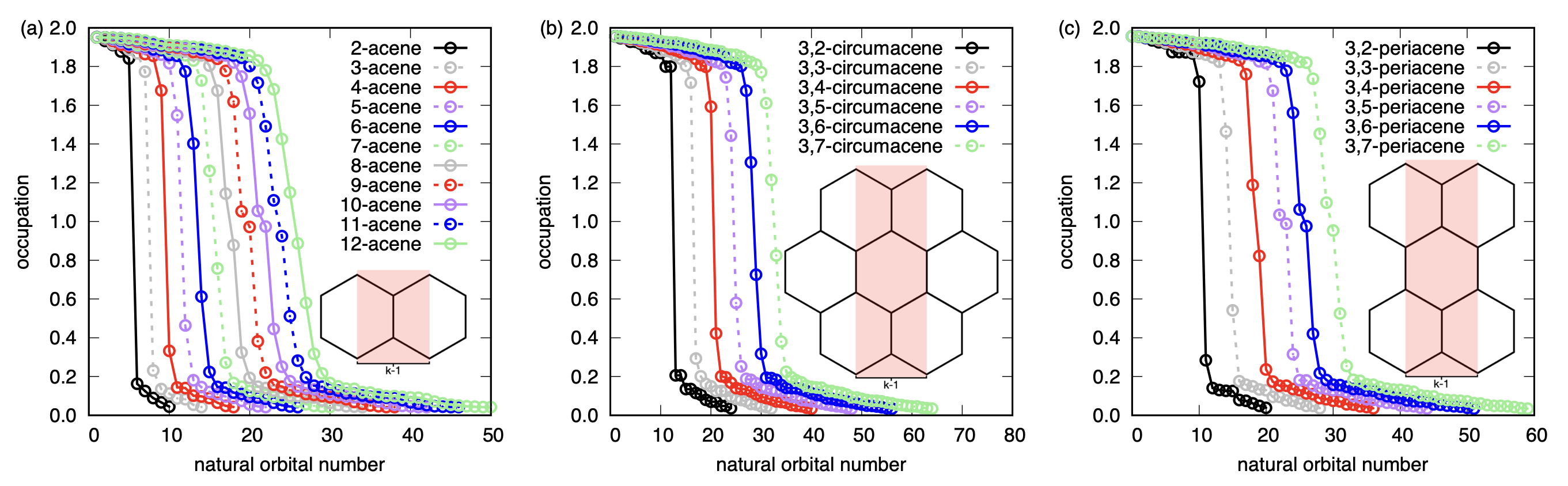}
    \end{center}
\end{figure*}

One common application of large-scale v2RDM calculations\cite{Mazziotti08_134108, Mazziotti11_5632, DePrince16_423, DePrince16_2260, DePrince19_290, DePrince19_276, DePrince19_6164, DePrince20_2274} has been the the electronic structure of zig-zag graphene nanoribbons,\cite{Han:2007:206805} a class of materials that are potential components in optoelectronic devices\cite{Bendikov:2004:4891,Anthony:2008:452,Yoo:2004:5427,Mayer:2004:6272,Chu:2005:243506,Paci:2006:16546,Rand:2007:659,Yang07_103501} and whose complex electronic structure has been the focus of many theoretical studies using both single- and multi-reference methods.
\cite{Nendel01_5517,Jiang:2008:332,Hajgato:2009:224321,Wu:2015:2003,Ibeji:2015:9849,Wudl04_7416,Chan07_134309,HeadGordon09_9779,Hajgato:2009:224321,Mazziotti08_134108,Yanai13_401,Evangelista16_161106,HeadGordon17_602,Leininger17_3746,HeadGordon18_547,Casula18_134112,DePrince16_2260,DePrince16_423,Mazziotti11_5632,Evangelista16_161106,Evangelista18_6295,Gagliardi17_2741,Gagliardi19_1716,DePrince19_290, DePrince19_6164, DePrince19_276,Lischka13_2581,Lischka14_1511}
One of the challenges in modeling these systems is that longer nanoribbons display complex multireference character that manifests in multiple partially occupied orbitals ({\em i.e.}, singlet polyradical character). Figure \ref{FIG:ACENES} depicts natural orbital occupation numbers for linear oligoacene molecules [k-acenes, panel (a)], 3,k-circumacene molecules [panel (b)], and 3,k-periacene molecules [panel(c)] as a function of the chain length, k. These data, taken from Ref.~\citenum{DePrince19_6164}, were computed using v2RDM-CASSCF and the cc-pVDZ basis set, with the active space chosen to consist of the valence $\pi$ and $\pi^*$ orbitals. {\color{black}The calculations enforced the PQG $N$-representability conditions.} Additional details regarding the specific orbitals within the active space and the geometries for the molecules can be found in Ref.~\citenum{DePrince19_6164}. For each class of molecules, we can see a transfer of density from the highest occupied natural orbital (HONO) to the lowest unoccupied natural orbital (LUNO) as the length of the molecule increases. For the k-acene and 3,k-periacene molecules, we see a crossing between the HONO and LUNO occupations, after which even more complicated polyradical character emerges. This polyradical character is most extreme for the 3,k-periacene molecules. Additional measures of polyradical character provided in Ref.~\citenum{DePrince19_6164} (effectively unpaired electrons\cite{Takatsuka:1978:175,Staroverov:2000:161,HeadGordon:2003:508,HeadGordon:2003:488} and changes in bond-length alternation patterns\cite{DePrince19_276}) confirm that the 3,k-periacene molecules display the most rapid and severe onset of polyradical character. Note that the active space for the largest system considered in Ref.~\citenum{DePrince19_6164}, 3,7-circumacene, was (64e, 64o). The corresponding v2RDM-CASSCF calculation required less than four hours of walltime when the code was executed on a combination of an NVIDIA TITAN V GPU and an Intel Core i7-6850k CPU.

In addition to the applications discussed above, large-scale v2RDM and v2RDM-CASSCF calculations have been applied in many other contexts. Several groups have studied hydrogen chains and clusters, which serve as both models for metal-insulator transitions and challenging benchmark materials for multi-reference correlation methods. For example, Sinitskiy, Greenman, and Mazziotti showed that v2RDM with PQG conditions can capture the transition between metal and insulating regimes in a $4\times 4\times 4$ cube of hydrogen atoms.\cite{Mazziotti10_014104} Stair and Evangelista\cite{Evangelista20_104108} applied v2RDM with PQG and PQG+T2 $N$-representability conditions to 1-, 2-, and 3-dimensional hydrogen clusters and found that PQG performed well for the 1-dimensional structures, but errors approaching 200 mE$_{\rm h}$ were obtained for 2- and 3-dimensional structures. The application of the T2 condition improves the situation, but errors approaching 50 mE$_{\rm h}$ were still observed. Li, Liebenthal, and DePrince\cite{DePrince21_174110} showed that the application of full 3POS improves the energetics for the 2- and 3-dimensional H$_{10}$ structures, but errors on the order of 20-30 mE$_{\rm h}$ were still observed. Aside from these model systems, Mazziotti and coworkers have applied v2RDM methods to a number of interesting systems, including conductive polymers,\cite{Mazziotti23_045001} nitrogonese cofactor,\cite{Mazziotti18_4988} cadmium telluride polymers,\cite{Mazziotti17_3142} and manganese porphyrin complexes.\cite{Mazziotti17_4656}

\subsection{Challenges for v2RDM Methods}

\subsubsection{Fractional Charges}

\begin{figure}[!htpb]
    \begin{center}
            \caption{Potential energy curves for the dissociation of a BH molecule when (a) enforcing and (b) neglecting spin-symmetry constraints.}
        \label{FIG:BH}
        \includegraphics{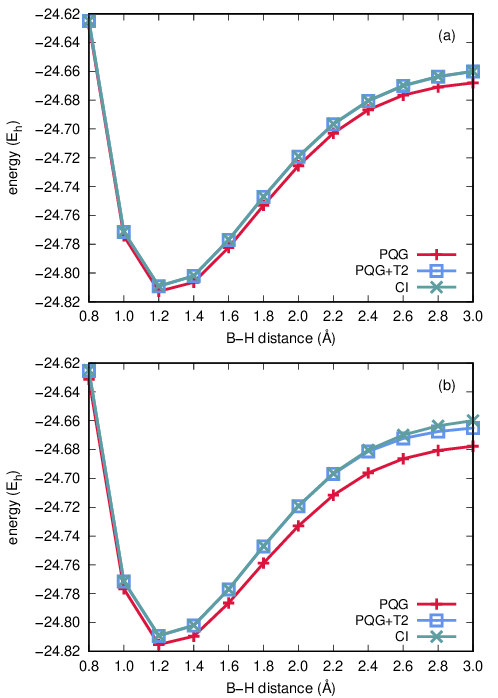}
    \end{center}
\end{figure}

In this subsection, we highlight some situations that pose challenges for v2RDM calculations performed under necessary but insufficient $N$-representability conditions. To begin, consider the dissociation of a simple molecule, boron hydride (BH), described within a minimal (STO-3G) basis set. Figure \ref{FIG:BH} provides potential energy curves for this molecule computed using full CI and v2RDM methods, using the PQG and PQG+T2 $N$-representability conditions. Panel (a) provides v2RDM curves computed when imposing typical spin-symmetry constraints, {\em i.e.}, $\hat{S}_z$ constraints (Eqs.~\ref{EQN:TR_D1}, \ref{EQN:TR_D2}, and \ref{EQN:D2_TO_D1}) and $\hat{S}^2$ constraints (Eq.~\ref{EQN:S2}), plus the maximal spin-projection constraints described in Ref.~\citenum{Ayers12_014110}. The curves in panel (b) were computed without spin-symmetry constraints. In this case, the trace constraints apply to the entire 1RDM and 2RDM (rather than individual spin blocks), and the contraction constraints involve partial traces over the entire 2RDM (again, rather than the individual spin blocks). In either case, it is clear that the behavior in the dissociation limit contrasts with the behavior in Fig.~\ref{FIG:N2_PEC}: different dissociation limits are reached by different methods. First, consider the cases where we impose spin-symmetry constraints [panel (a)]. We can see that PQG significantly underestimates the energy at a B--H distance of 3 \AA~(by roughly 8 $\times 10^{-3}$ E$_{\rm h}$), while PQG+T2 gives results that are indistinguishable from CI on the scale of the figure. Without spin-symmetry constraints, PQG and PQG+T2 both underestimates the energy at 3 \AA~by 18 $\times 10^{-3}$ E$_{\rm h}$ and 5 $\times 10^{-3}$ E$_{\rm h}$, respectively. The reason the wrong dissociation limit is reached is because, at these levels of theory, it is energetically favorable for the molecule to dissociate into fractionally charged species, rather than neutral boron and hydrogen atoms. 

\begin{figure}[!htpb]
    \begin{center}
            \caption{Energy of a boron atom as a function of the number of electrons.}
        \label{FIG:FRACTIONAL_CHARGES}
        \includegraphics{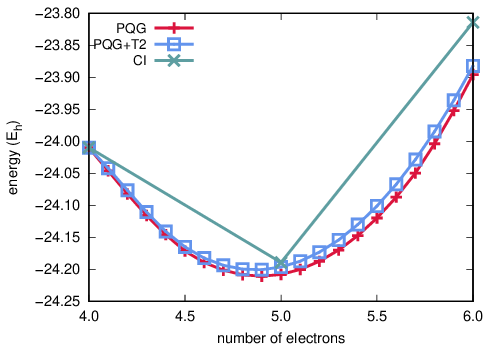}
    \end{center}
\end{figure}

The failure of v2RDM to describe the dissociation limit in BH is reminiscent of the fractional charge problem in density functional theory (DFT) when using approximate density functional approximations (DFAs), which is related to the lack of a discontinuity of the energy of a DFA as a function of fractional numbers of particles.\cite{Yang08_792} Indeed, v2RDM theory suffers from the same issue.\cite{Ayers09_5558, Bultinck10_114113} Figure \ref{FIG:FRACTIONAL_CHARGES} shows the energy of a boron atom as a function of the number of electrons, as predicted by v2RDM calculations performed under the PQG and PQG+T2 conditions, when using the STO-3G basis set. Note that, for fractional numbers of electrons, constraints on $\hat{S}_z$ and $\hat{S}^2$ do not make sense, so these calculations were performed in the absense of such spin symmetry constraints [like those from Fig.~\ref{FIG:BH}(b)]. As one might expect, given the incorrect dissociation limits reached in Fig.~\ref{FIG:BH}, both the PQG- and PQG+T2-derived energies for boron are convex functions of the number of electrons, without any derivative discontinuity at a value of five electrons (the neutral species). For both methods, the energy is lowest for a value of 4.9 electrons, by about 2.8 $\times 10^{-3}$ E$_{\rm h}$ and 4.1 $\times 10^{-3}$ E$_{\rm h}$ for PQG and PQG+T2, respectively. Unfortunately, there is no guarantee that the application of higher-order constraints ({\em e.g.}, 3POS) will remedy this situation, until the complete set of $N$-representability conditions are reached. This sort of an analysis paints a grim picture for v2RDM methods, but the situation is not quite as bleak as Fig.~\ref{FIG:FRACTIONAL_CHARGES} might suggest. For example, PQG+T2 lacks a derivative discontinuity, but the application of spin-symmetry constraints leads to essentially exact agreement with CI, for this system (Fig.~\ref{FIG:BH}). On the other hand, the data in Fig.~\ref{FIG:BH} show that the wrong dissociation limit is reached for PQG regardless of our spin symmetry considerations. Note that Verstichel, van Aggelen, Van Neck, Ayers, and Bultinck\cite{Bultinck10_114113} have attempted to resolve this issue through the application of subsystem constraints based on fractional-$N$ ensemble $N$-representability. While this approach leads to the correct dissociation limit for diatomic molecules, its application to general polyatomic systems would be challenging because it would require multiple calculations on many subsystems (constituent atoms, pairs of atoms, trimers of atoms, etc.), possibly with multiple charge states. 

\subsubsection{Angular Momentum States}

For a molecule described by a non-relativistic Hamiltonian, in the absence of a magnetic field, the energies associated with states having a given total spin quantum number, $S$, and various projection-of-spin quantum numbers, $M_S$, should be degenerate. However, v2RDM methods fail to respect this degeneracy.  An excellent discussion these issues can be found in Ref.~\citenum{Ayers12_014110}. In that work, van Aggelen, Verstichel, Bultinck, Van Neck, and Ayers develop a number of ensemble and pure spin state constraints and discuss the representation of the 1RDM, 2RDM, etc. using spin-symmetry-adapted basis functions (see also Gidofalvi and Mazziotti's discussion of spin-adapted basis functions in Ref.~\citenum{Mazziotti05_052505}). Under these spin-symmetry constraints, one finds that maximal spin projection states having $|M_S| = S$ are better constrained (meaning that they have a higher energy) than spin projection states with $|M_S| < S$. Given this behavior, calculations on non-singlet molecules are typically carried out for the maximal spin-projection state. 

For systems that possess additional angular momentum symmetries, one could could develop expectation-value-type constraints on the relevant operators. For example, Li and DePrince\cite{DePrince19_032509} have considered constraints on the orbital angular momentum
\begin{equation}
\label{EQN:L2}
\langle \hat{L} \rangle = L(L+1)
\end{equation} 
and on the $z$-projection of the orbital angular momentum
\begin{align}
    \label{EQN:LZ}
    \langle \hat{L}_z \rangle &= M_L \\ 
    \label{EQN:LZ2}
    \langle \hat{L}^2_z \rangle - \langle \hat{L}_z \rangle^2 &= 0
\end{align}
Equations \ref{EQN:L2}, \ref{EQN:LZ} and \ref{EQN:LZ2} are relevant in atomic systems, while Eqs.~\ref{EQN:LZ} and \ref{EQN:LZ2} can be applied to calculations on linear molecules. Reference \citenum{DePrince19_032509} shows that, for states with non-zero $L$, these constraints can have non-trivial energetic impacts ($> 10$ mE$_{\rm h}$ for some second-row atoms). As for linear molecules, $M_L$ constraints can also have significant impacts in certain cases. For example, in the absence of any orbital angular momentum symmetry considerations, v2RDM calculations incorrectly predict that the ground state of molecular oxygen, described by the cc-pVDZ basis set, is a singlet; constraints on the expectation value and variance of $\hat{L}_z$ remedy this issue. Aside from these successes, v2RDM fails to describe states with a given $L$ and various $M_L$ values in a balanced way. As we discussed in the context of spin symmetry, for a given $L$, states with maximal orbital angular momentum projection ($|M_L| = L$) are more well constrained than those with $|M_L| < L$.  

Given the issues describing various spin and orbital angular momentum projection states, it is not surprising that a similar issue persists in the case of total angular momentum. Reference \citenum{DePrince22_5966} considered total angular momentum constraints  of the form
\begin{align}
    \label{EQN:J2}
    \langle \hat{J} \rangle &= J(J+1) \\
    \label{EQN:JZ}
    \langle \hat{J}_z \rangle &= M_J \\ 
    \label{EQN:JZ2}
    \langle \hat{J}^2_z \rangle - \langle \hat{J}_z \rangle^2 &= 0
\end{align}
with $\hat{J} = \hat{L} + \hat{S}$ and $\hat{J}_z = \hat{L}_z + \hat{S}_z$.
When enforcing these constraints in calculations on atomic systems, v2RDM fails to maintain the degeneracy of the $J$ states, for a given $S$ and $L$, and, given fixed $L$, $S$, and $J$, the $M_J$ states are not degenerate either. These failures in the non-relativistic limit suggest that, in the absence of additional $N$-representability constraints, v2RDM calculations cannot be applied with any expectation of reliability in the context of relativistic calculations. 

\subsubsection{Pure-State Versus Ensemble-State $N$-representability}

This review article has focused on ensemble-state $N$-representability constraints. For correlated RDMs, the pure-state problem is far more challenging and involves complex and non-intuitive inequality constraints, known as generalized Pauli constraints (GPCs),\cite{Klyachko06_72, Klyachko08_287} on the eigenvalues of one-body RDMs. These conditions were discovered empirically in the 1970s (for a system of six electrons in six spin orbitals),\cite{Dennis72_7} and it was not until the mid 2000s that an automated procedure was developed for the generation of GPCs in other systems.\cite{Klyachko06_72,Klyachko08_287} The number of GPCs grows quickly; for only five electrons in ten orbitals there are already 160 conditions, and there are a combinatoral number of GPCs, in general. 
{\color{black}For even numbers of electrons, the GPCs on the 1RDM reduce to only a requirement of pairwise degeneracy of the natural orbital occupation numbers. In such cases, the GPCs are trivially enforced in v2RDM and other density matrix functional theories.\cite{Pernal19_012509} Given the complexity of the general (odd electron number) case, however, few} papers describe protocols to enforce GPCs in the context of v2RDM-based optimizations.\cite{DePrince16_164109, Mazziotti19_022517} For a discussion of how GPC constraints can be applied to the 2RDM, see Refs.~\citenum{DePrince22_5966} and \citenum{Mazziotti16_032516}.

\section{Outlook}

\label{SEC:CONCLUSIONS}

The most successful application of the v2RDM approach has been its use in the context of large-active-space CASSCF calculations.\cite{Mazziotti08_134108, Mazziotti11_5632, DePrince16_423, DePrince16_2260, DePrince19_276, DePrince19_6164} With two-particle (PQG) $N$-representability conditions, active spaces as large as (64e, 64o) can be treated, with a suitable algorithm, in a matter of only a few hours.\cite{DePrince19_6164} However, one can only expect such calculations to offer qualitative insights into the electronic structure. Quantitatively accurate v2RDM-CASSCF calculations would require the application of partial three-particle conditions ({\em e.g.}, T2) or full 3POS, in which case the active space sizes that could reasonably be treated would be much smaller. Still, Mazziotti has reached active space sizes of (30e, 30o) when enforcing the T2 condition using his low-rank dual cone optimization algorithm.\cite{Mazziotti16_153001,Mazziotti20_052819} Other interesting use cases include alternative schemes for capturing strong correlation effects, such as v2RDM-based approximations to DOCI.\cite{VanNeck15_4064, Mazziotti17_084101, DeBaerdemacker18_024105, Lain18_194105, DePrince19_244121, Alcoba21_013110} In either case, one must contend with missing dynamical correlation, which can be done with standard post-CASSCF (or DOCI) methods, such as multiconfiguration pair density functional theory\cite{DePrince19_290, DePrince20_2274, Massaccesi20_e26256} or multireference adiabatic connection techniques.\cite{DePrince19_244121, DePrince20_4351}

The machinery of v2RDM is also potentially useful in the context of error mitigation for quantum chemical calculations on noisy intermediate scale quantum (NISQ) devices. The basic premise is, given a noisy RDM measured on a NISQ era device, a v2RDM-like procedure could be used to find the closest possible 2RDM that satisfies important $N$-representability conditions.
Several examples of error mitigation and variance reduction strategies built upon both ensemble-state\cite{Mazziotti12_012512,McClean18_053020,Mazziotti19_022517,Mazziotti21_012420,DiCarlo19_010302,Miyake20_arxiv,Chan23_2306.05640} and pure-state\cite{Mazziotti19_022517,Zalcman20_6507} $N$-representability can be found in the literature. This use case is less well explored than the direct application of v2RDM in the context of (classical) quantum chemical simulations.

Lastly, note that, as described in this review, the v2RDM approach is only directly applicable to the ground electronic state of a given spin or orbital angular momentum symmetry. The reason for this limitation is that we lack general constraints that can differentiate RDMs for ground and excited states (with the exception of expectation-value-type constraints applied to angular momentum operators, for example). Excited states can be extracted from post v2RDM procedures such as the extended random phase approximation (ERPA),\cite{VanNeck13_50, DePrince16_164109, DePrince18_234101} but the ERPA is of limited utility, in general, because it assumes that the excited states are of single excitation character. Moreover, ERPA excitation energies can be quite poor when the ground-state 1RDM is not pure-state $N$-representable.\cite{VanNeck13_50} The application of pure-state conditions fixes this issue,\cite{DePrince16_164109} but the resulting v2RDM optimization is far more computationally demanding than the ensemble-state optimization, and pure-state conditions on the 1RDM are only known for small systems. Given these issues, the description of excited states in the context of v2RDM is an open challenge.

\vspace{0.5cm}


\vspace{0.5cm}

\begin{acknowledgments}This material is based upon work supported by the National Science Foundation under Grant No. CHE-2100984.\\ 
\end{acknowledgments}



\bibliography{master.bib}

\end{document}